\documentclass[twocolumn,aps,prb,floats]{revtex4}
\usepackage{graphicx}
\usepackage{amsmath}
\newcommand{\half}{\frac 12}
\newcommand{\sgn}{\mathop{\mathrm{sgn}}}
\newcommand{\sigmab}{\overline \sigma}
\newcommand{\ef}{\varepsilon_F}
\begin{document}
\title{Effect of Strong Correlations on the Disorder-Induced Zero Bias Anomaly in the Two-Site Anderson-Hubbard Model}
\author{Hong-Yi Chen}
\affiliation{Department of Physics, National Taiwan Normal University, Taipei 11677, Taiwan}
\author{W. A. Atkinson} \email{billatkinson@trentu.ca}
\affiliation{Department of Physics and Astronomy, Trent University, 1600 West Bank Dr., Peterborough ON, K9J 7B8, Canada}
\begin{abstract}
Several recent exact diagonalization calculations have established that the Anderson-Hubbard model has a disorder-induced zero bias anomaly (ZBA) (also called a disorder-induced pseudogap) in the density of states.  In order to understand the physics of 
the ZBA, we study a simplified problem---an ensemble of two-site molecules with random site energies---for which analytical results are possible.
For this ensemble, we examine how the ZBA forms in both the weakly correlated (mean field) and strongly correlated limits.   In the weakly correlated case, the ZBA can be understood as the result of level repulsion between bonding and antibonding molecular orbitals.  A similar level repulsion occurs in the strongly correlated case too, but a larger contribution to the ZBA comes from the suppression of a triplet excitation mode.  This inherently many-body mechanism does not have a counterpart in mean-field models.
\end{abstract}
\maketitle

\section{Introduction}
A number of recent papers have shown the existence of a disorder-induced zero bias anomaly (ZBA)  in the Anderson-Hubbard model (AHM) in one and two dimensions.\cite{Song2009,Chiesa2008,Shinaoka2009,Shinaoka2009b,Hongyi2010a} These calculations have revealed that there is a V-shaped dip in the density of states at the Fermi energy $\varepsilon_F$.  This dip is  produced by the response of the inelastic self-energy to the disorder potential.\cite{Song2009,Hongyi2010a}  Such a mechanism is well-understood in conventional metals and insulators, where the effect was explained at the level of Hartree-Fock theory by Altshuler and Aronov.\cite{Altshuler1985}  However, strong correlation effects are generally important  in the AHM, and the Altshuler-Aronov mechanism is thus insufficient for this case.\cite{Hongyi2010a}

The AHM is the standard model for strongly-correlated systems with disorder.  Like the Hubbard model, electrons are assumed to move on a tight-binding lattice of atomic-like orbitals.  A zero-range intraorbital Coulomb interaction $U$ is included, but longer range interorbital interactions are neglected.  Strong correlations are important when the intersite hopping matrix element $t$ is small relative to $U$.  The AHM differs from the Hubbard model by the addition of disorder, which is introduced by selecting the orbital energies $\epsilon_i$ from a random distribution of width $\Delta$ (the subscript $i$ labels sites in the atomic lattice).  The Hamiltonian is
\begin{equation}
H = -t \sum_{\langle ij \rangle,\sigma } c^{\dagger}_{i\sigma} c_{j\sigma} +\sum_i \left (
\epsilon_i \hat n_i + U \hat n_{i\uparrow} \hat n_{i\downarrow} \right )
\end{equation}
where 
$\langle ij\rangle$ restricts the sum to nearest neighbor sites, $\hat n_{i\sigma}$ is the number operator for site $i$ and spin $\sigma$, $\hat n_{i} = \sum_{\sigma} \hat n_{i\sigma}$, and $\epsilon_i \in [-\half \Delta, \half \Delta]$.
(We use $\hat O$ to indicate the operator form of an observable O; thus $n_i = \langle \hat n_i \rangle$.)
In this model, the ensemble-averaged density is $n=1$ (i.e.\ the band is half-filled) for $\ef = \frac U 2$.

\begin{figure}
\includegraphics[width=\columnwidth]{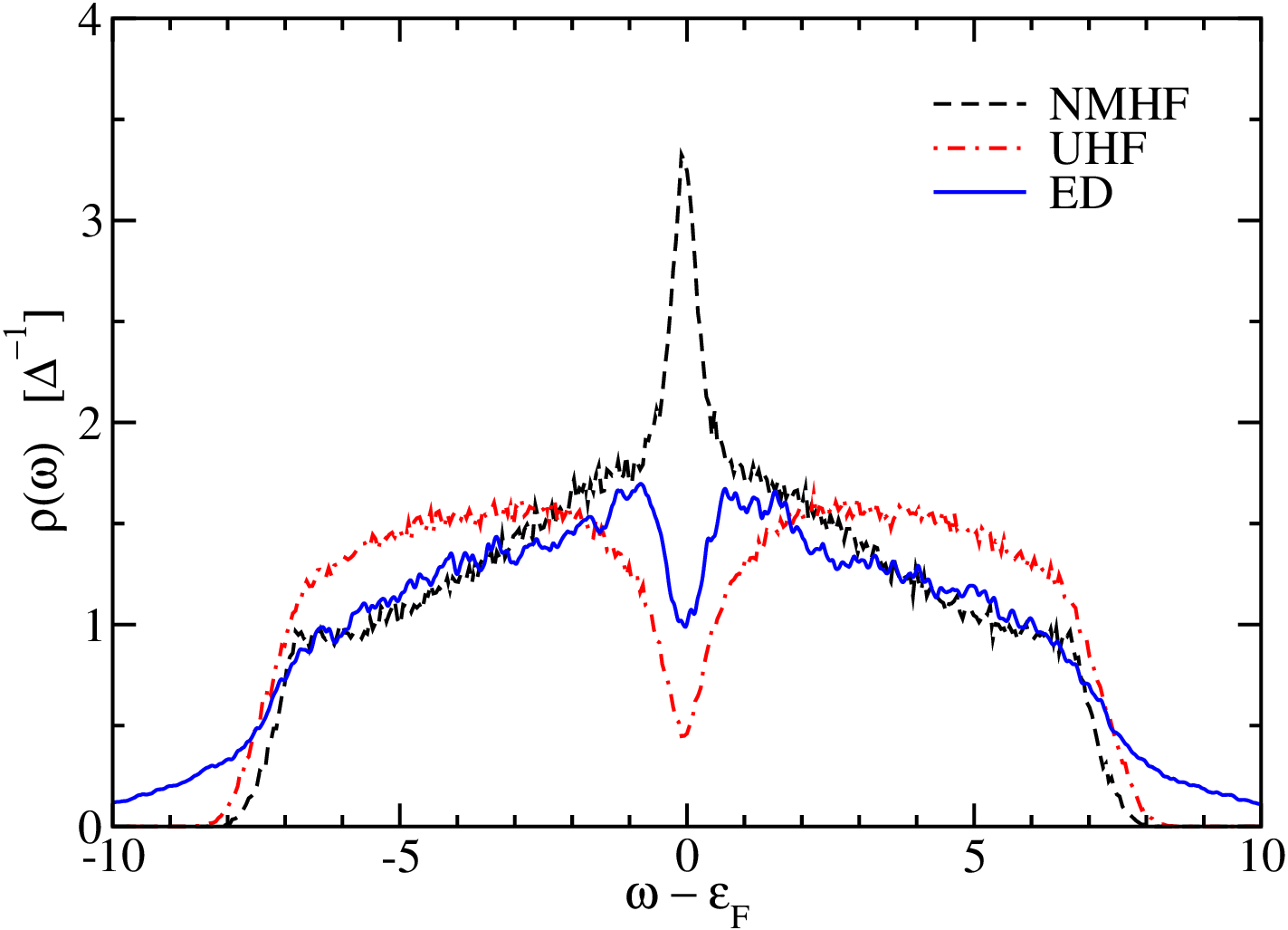}
\caption{(color online) Comparison of densities of states for the Anderson-Hubbard model using different approximations.  Results are shown for the nonmagnetic Hartree-Fock (NMHF), unrestricted Hartree-Fock (UHF) and exact diagonalization (ED) of small clusters.  Hartree-Fock calculations are ensemble-averaged self-consistent calculations for $10\times 10$ lattices and 1000 impurity configurations.  Exact diagonalization calculations are for a 12-site lattice and 1000 impurity configurations.  Model parameters are $\Delta=20$, $U=8$, $t=1$ and $\ef=U/2$, corresponding to half-filling.}
\label{fig:cmpr_dos}
\end{figure}

The conventional Altshuler-Aronov theory predicts that the Hartree and
exchange self-energies make positive and negative contributions to the
density of states at $\ef$ respectively.\cite{Altshuler1985} The
exchange self-energy is typically much larger than the Hartree
self-energy, and the net result is a depletion of states at $\ef$.
However, the AHM has a zero-range interaction for which the exchange
self-energy vanishes.  Altshuler-Aronov theory predicts a peak in this
case, which is illustrated by the nonmagnetic Hartree-Fock calculations in
Fig.~\ref{fig:cmpr_dos}.
%
This is in contrast to the V-shaped dip found in exact diagonalization
calculations.\cite{Chiesa2008,Shinaoka2009,Shinaoka2009b} The
Altshuler-Aronov prediction assumes a nonmagnetic ground state, and a
number of unrestricted Hartree-Fock calculations have found a V-shaped
dip at
$\varepsilon_F$\cite{Tusch1993,Fazileh2006,ChenXi2009,Shinaoka2010} in
the magnetic phase.\cite{ChenXi2009} While the unrestricted
Hartree-Fock results are qualitatively similar to the exact
diagonalization results, there are some important differences.
Notably, the ZBA in the unrestricted Hartree-Fock calculations grows
with increasing $U$, eventually forming a broad soft gap when $U$ is
sufficiently large.  In contrast, the ZBA in exact diagonalization
calculations saturates for large $U$ (provided $U<\Delta$; a Mott gap
opens for $U\gtrsim\Delta$), and empirically has a width $\sim
t$.\cite{Chiesa2008,Hongyi2010a} Densities of states based on the
different approximations are illustrated in Fig.~\ref{fig:cmpr_dos}.

We note that the above discussion ignores the low energy soft
gap\cite{Shinaoka2009,Shinaoka2009b,Shinaoka2010} that has been
inferred from exact diagonalization in one dimension, and found in
unrestricted Hartree-Fock calculations in one and three dimensions.
This gap appears on a scale $|\omega-\ef| \lesssim O(0.1t)$, and has
been ascribed to long range
correlations.\cite{Shinaoka2009,Shinaoka2009b} The current work
examines the two-site AHM where long range correlations are absent,
and there is no soft gap.

\begin{figure}
\includegraphics[width=\columnwidth]{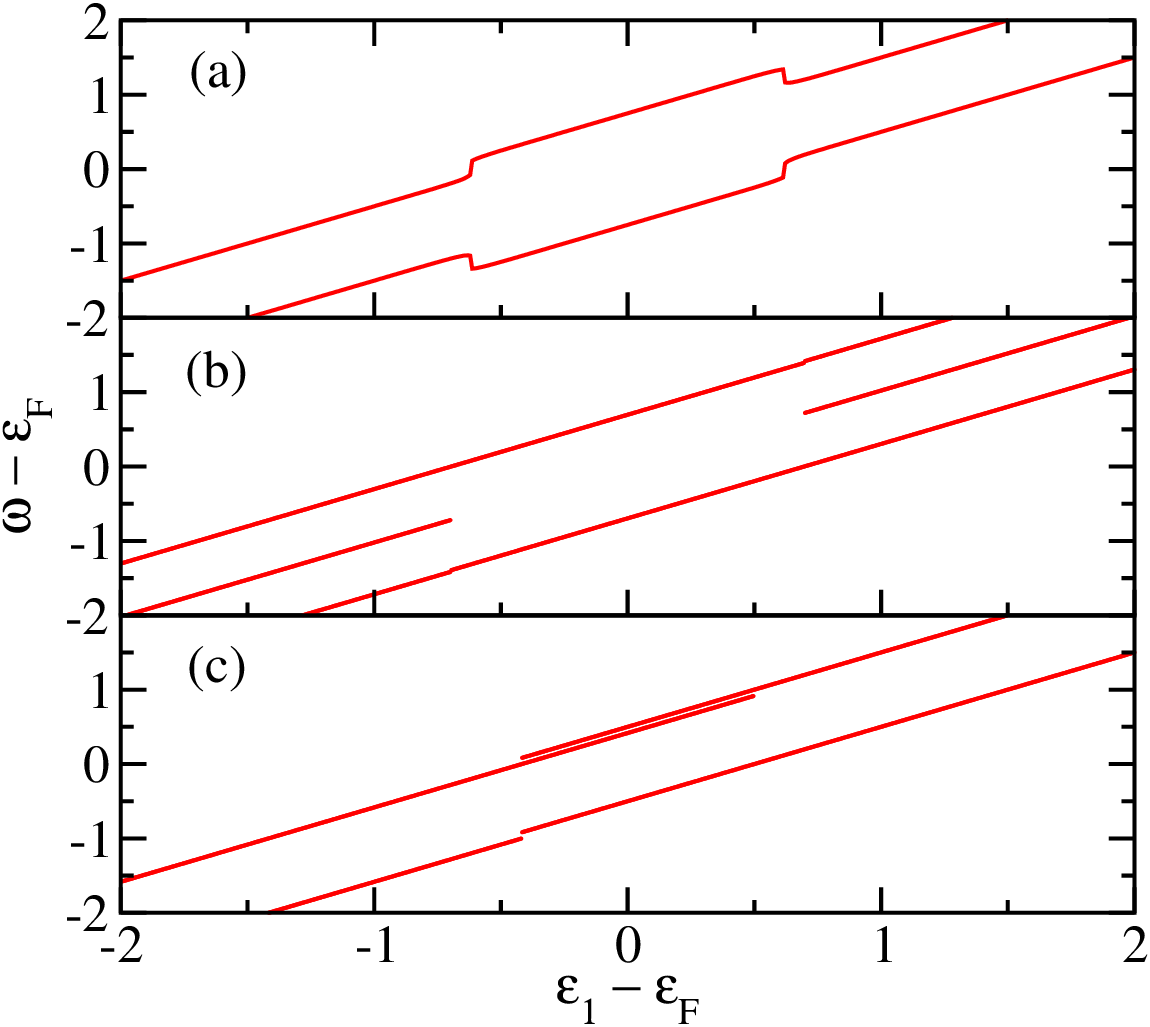}
\caption{(color online) Low energy excitation spectra as a function of site energy for the degenerate two-site model.  Lines represent peak position $\omega$ in the tunneling density of states, plotted as a function of $\epsilon_1$ for $t=0.5$.  (a) Excitation spectrum for the mean-field Hamiltonian (\ref{eq:H2}) with $V=0.5$ and $\epsilon_2=\epsilon_1$; (b) Excitation spectrum for exact-diagonalisation calculations with $\epsilon_2 = \epsilon_1-U$ and (c) $\epsilon_2=\epsilon_1$.  Both (b) and (c) have $U=12$.}
\label{fig:levelsplit}
\end{figure}

The advantage of the two-site AHM is that it is simple enough that analytical results are possible, and yet is sufficiently rich to explain much of the physics of the ZBA in larger systems.\cite{Wortis2010}  
Here, our goal is to compare the two-site AHM to a simple mean-field two-site model in order to answer the question: in what way is the ZBA in strongly correlated systems different from that in
conventional metals?  

%
%
Our main results are summarized in Fig.~\ref{fig:levelsplit}.  In Fig.~\ref{fig:levelsplit}(a), we plot the low-energy excitation spectrum for a pair of sites with energies $\epsilon_1=\epsilon_2$ using the mean-field model described in Sec.~\ref{HF}.  This model is meant to illustrate the conventional Altshuler-Aronov mechanism for the ZBA.  Without interactions, the hybridization of atomic orbitals leads to a level splitting of $2t$ between bonding and antibonding orbitals.  With interactions, there is a range of site energies near $\ef$ where the level splitting is larger than $2t$.  This enhancement of level splitting (i.e.\ this level repulsion) shifts spectral weight away from $\ef$ and is the origin of the ZBA in this model.
In Fig.~\ref{fig:levelsplit}(b), an alternative mechanism for shifting spectral weight away from $\ef$ is presented.  In this case, exact results for the low-energy excitation spectrum of the two-site AHM are shown.  We have taken $\epsilon_1=\epsilon_2+U$, which means that the lower Hubbard orbital  of site 1 is degenerate with the upper Hubbard orbital of site 2.  Here, the spectrum has three excitation poles, the middle of which is a triplet excitation.  The gap which is evident in the triplet spectrum is one of the main reasons for the pronounced ZBA in the two-site AHM, and is an inherently many-body mechanism that lies outside the mean-field Altshuler-Aronov paradigm.  Finally, in Fig.~\ref{fig:levelsplit}(c), we show that interactions have little effect on the spectrum if we consider the case of degenerate orbitals $\epsilon_1 = \epsilon_2$.   In this case, excitation spectra are shifted by $\sim O(t^2/U)$ from their noninteracting values.  

The goal of this paper is to explore the physics behind these results.  We discuss the mean-field mechanism for the ZBA in Sec.~\ref{HF}, and emphasize in particular the role of level repulsion.  We then derive, in Sec.~\ref{strong}, an expression for the ensemble-averaged density of states for the two-site AHM.  Finally, we discuss in Sec.~\ref{discussion} the different mechanisms by which the ZBA found in Sec.~\ref{strong} arises.   

\section{ZBA in Mean-Field Theory}
\label{HF}
It is worth reviewing briefly how the ZBA arises in conventional metals.  A variety of physical explanations for the Altshuler-Aronov ZBA have been given,\cite{Altshuler1985,Abrahams1981,Rudin1997} and in this work we adopt the language of level repulsion.\cite{Levit1999}  

We consider an ensemble of two-site AHMs with randomly chosen site energies.
Since we restrict ourselves to nonmagnetic solutions of the Hartree-Fock equations, a  V-shaped ZBA is possible only if a nonlocal interaction is included.  We therefore add a repulsive interaction $V\hat n_{1}\hat n_2$ to the Hamiltonian.  In Hartree-Fock theory,
\begin{eqnarray}
V\hat n_1\hat n_2 &\rightarrow& V \left ( n_1\hat n_2 + n_2\hat n_1 \right ) \nonumber \\
&& - V \sum_\sigma \left (
\langle c_{1\sigma}^\dagger c_{2\sigma}\rangle c_{2\sigma}^\dagger c_{1\sigma} + 
h.c.
\right ).
\end{eqnarray}
The first and second terms are the Hartree and exchange contributions respectively, and there is an additional Hartree contribution $\frac 12 U\sum_i n_{i}\hat n_{i} $ from the on-site interaction.  
The Hartree contribution to the density of states is small for weak disorder\cite{Altshuler1985} but is central to the physics of the Coulomb gap for large disorder; the exchange contribution is largest for weak disorder, and underlies the Altshuler-Aronov mechanism for the ZBA.  Because our goal is to contrast the Altshuler-Aronov mechanism with the physics of the the AHM, we discuss only the exchange term.   

\begin{figure}[tb]
\includegraphics[width=\columnwidth]{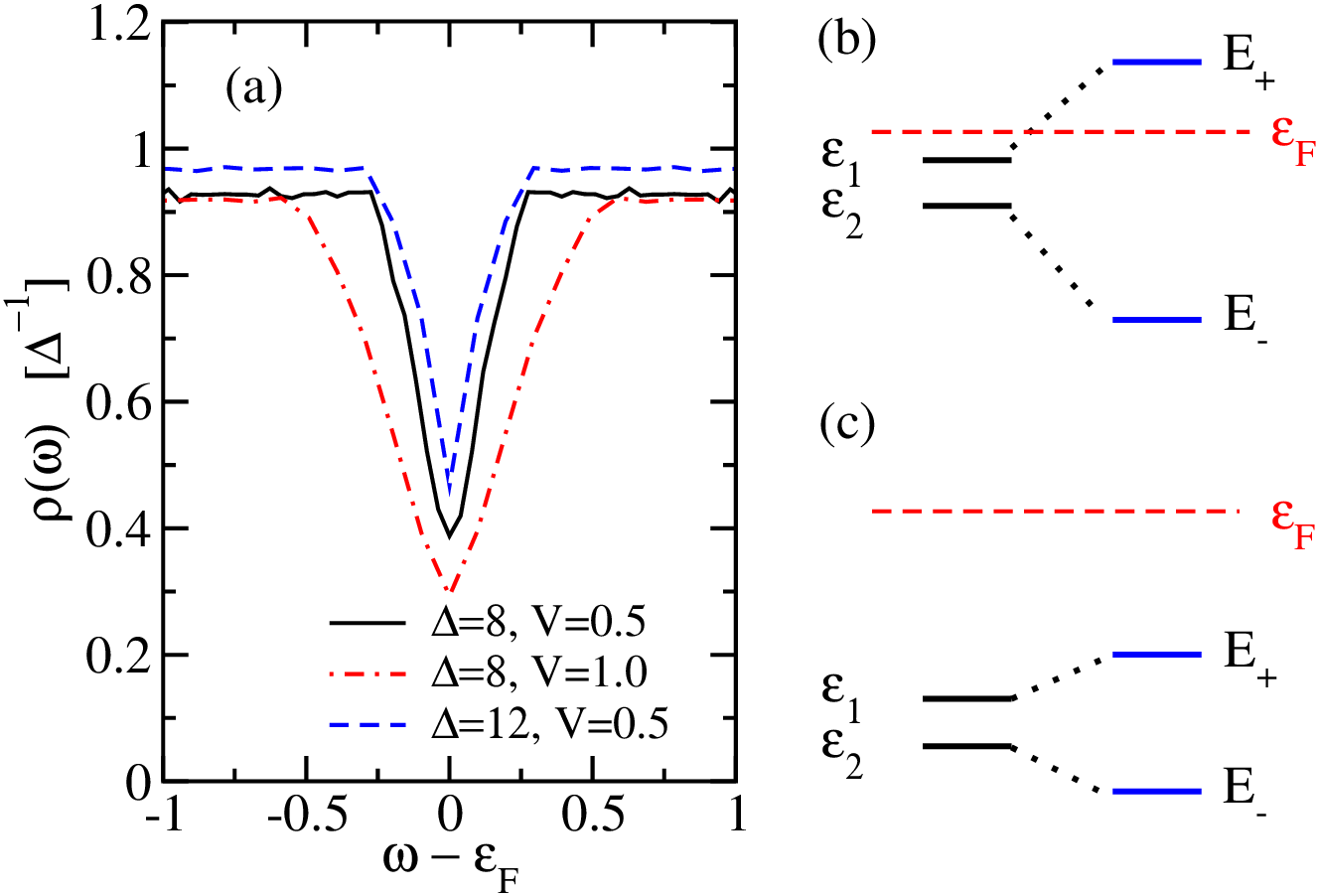}
\caption{(color online) Zero bias anomaly for self-consistent solutions of the mean-field Hamiltonian (\protect\ref{eq:H2}).  (a) The density of states for $t=1$ and different $\Delta$ and $V$. (b) The exchange self-energy is nonzero for configurations of $\epsilon_1$ and $\epsilon_2$ for which $E_+ > \ef > E_-$.  This leads to an enhanced level repulsion relative to configurations, as in (c) where $E_+$ and $E_-$ are on the same side of $\ef$ and the exchange self-energy vanishes.}
\label{fig:two-siteX}
\end{figure}

Neglecting the Hartree contributions, we obtain the
mean-field exchange Hamiltonian
\begin{equation}
H_X = \sum_{i} \epsilon_i \hat n_i - \tilde t \sum_\sigma \left( c_{1\sigma}^\dagger c_{2\sigma}
+ c_{2\sigma}^\dagger c_{1\sigma} \right)
\label{eq:H2}
\end{equation}
where the renormalized hopping matrix element is $\tilde t = t + V\langle c_{1\sigma}^\dagger c_{2\sigma}\rangle$.  The eigenergies of $H_X$ are 
\begin{equation}
E_{X,\pm} = \frac{\epsilon_1+\epsilon_2}{2} \pm \sqrt{ \left (\frac{\epsilon_1-\epsilon_2}{2}\right )^2 + \tilde t^2 },
\label{eq:epm}
\end{equation}
and a straightforward calculation yields
\begin{equation}
\langle c_{1\sigma}^\dagger c_{2\sigma}\rangle = -\tilde t\frac{f(E_{X,+})-f(E_{X,-})}{E_{X,+}-E_{X,-}}.
\label{eq:c1c2}
\end{equation}
Equations (\ref{eq:epm}) and (\ref{eq:c1c2}) allow $\tilde t$ to be determined self-consistently for each $(\epsilon_1,\epsilon_2)$ pair.  The ensemble-averaged density of states for this model exhibits a ZBA, as shown in Fig.~\ref{fig:two-siteX}.

The term ``level repulsion'' refers to the fact that the level spacing between molecular eigenenergies is greater than the level spacing between the original atomic energies, namely $E_{X,+} - E_{X,-} > |\epsilon_1 - \epsilon_2|$.  For a repulsive interaction $V$, $\tilde t > t$ and the level repulsion is enhanced by the exchange self-energy.  This enhanced level repulsion, by itself, does not lead to a dip in the density of states; it is necessary that the amount of level repulsion depend on the values of $E_{X,\pm}$ relative to $\varepsilon_F$. At zero temperature, Eq.~(\ref{eq:c1c2}) shows that $\tilde t$ is different from $t$ only if $E_{X,+} >\varepsilon_F>E_{X,-}$, as illustrated schematically in Fig.~\ref{fig:two-siteX}.  This has the effect of pushing states away from $\ef$, as shown numerically for the case $\epsilon_1=\epsilon_2$ in Fig.~\ref{fig:levelsplit}(a).  In this language, the ZBA in conventional metals is understood as level repulsion between filled and empty molecular orbitals near $\varepsilon_F$. 

\section{Approximate Diagonalization of the Two-Site AHM}
We now turn to an approximate solution of the two-site AHM that preserves strong-correlation physics. 
We work in the strongly-correlated limit $U \gg t$, where we can
isolate terms that contribute to the density of states on the energy
scale $t$.  Higher order terms, which contribute on the scale $t^2/U$,
are neglected.
We begin with a brief review of the atomic limit ($t=0$), where interactions already have a nontrivial effect on the density of states, and then show how the density of states is modified by a nonzero $t$.

\label{strong}
\subsection{Atomic Limit}
\label{atomic}

The density of states can be found exactly in the atomic limit $t=0$.  Each site is independent, and the ground state $|G_i\rangle$ for the $i$th site is 
\begin{equation}
|G_i\rangle = \left \{ \begin{array}{ll} |0\rangle, &  \ef<\epsilon_i \\
|\uparrow\rangle, & \ef-U < \epsilon_i < \ef \\
|2\rangle, &\epsilon_i<\ef-U
\end{array}\right .
\label{eq:1egs}
\end{equation}
We have assumed a weak Zeeman splitting so that spin-up states are preferred when there is  an odd number of electrons.  The spin-averaged retarded Green's function for the $i$th site is
\begin{eqnarray}
G_{i}(\omega) &=& \half \sum_{m\sigma} \left [ \frac{|\langle m|c_{i\sigma} |G_i\rangle|^2}{\omega^+-E_{G_i}+E_m}
+
 \frac{|\langle m|c_{i\sigma}^\dagger |G_i\rangle|^2}{\omega^++E_{G_i}-E_m} \right ]
\nonumber \\
&=& \frac{1-\half n_i}{\omega^+ -\epsilon_i} + 
\frac{\half n_{i}}{\omega^+ -\epsilon_i -U}
\end{eqnarray}
where $\omega^+ = \omega+i0$, $n_{i} = \sum_\sigma \langle \hat n_{i\sigma}\rangle$,  $|m\rangle$ are a complete set of excited states with energies $E_m$, and $E_{G_i}$ is the ground state energy.  The spin-averaged density of states at site $i$ is thus
\begin{eqnarray}
\rho_{\epsilon_i}(\omega) &=& -\frac{1}{\pi}\mbox{Im } G_{i}(\omega) \\
&=& \left( 1-\frac {n_i}{2}\right ) \delta(\omega-\epsilon_i) + \frac{n_{i}}{2} \delta(\omega-\epsilon_i -U).
\label{eq:1sdos}
\end{eqnarray}
This equation shows that (i) strong correlations split the local spectrum at each site into a pair of poles at $\epsilon_i$ and $\epsilon_i+U$ and (ii) the weight of each pole depends on the electron density at that site.  We refer to the poles at $\epsilon_i$ and $\epsilon_i+U$ as the lower Hubbard orbitals (LHO) and upper Hubbard orbitals (UHO) respectively.  
It is worth emphasizing that the energies of the LHO and UHO determine the total charge density at each site.  From Eq.~(\ref{eq:1egs}), 
\begin{equation}
n_{i} = \left \{ \begin{array}{lr}
0, & \varepsilon_F < \epsilon_i \, (\mbox{LHO and UHO above $\ef$}) \\
1, & \epsilon_i <  \varepsilon_F  < \epsilon_i + U  \, (\mbox{LHO below; UHO above}) \\
2, & \epsilon_i  + U < \varepsilon_F \, (\mbox{LHO and UHO below $\ef$})
\end{array}
\right . 
\end{equation}

\begin{figure}[tb]
\includegraphics[width=\columnwidth]{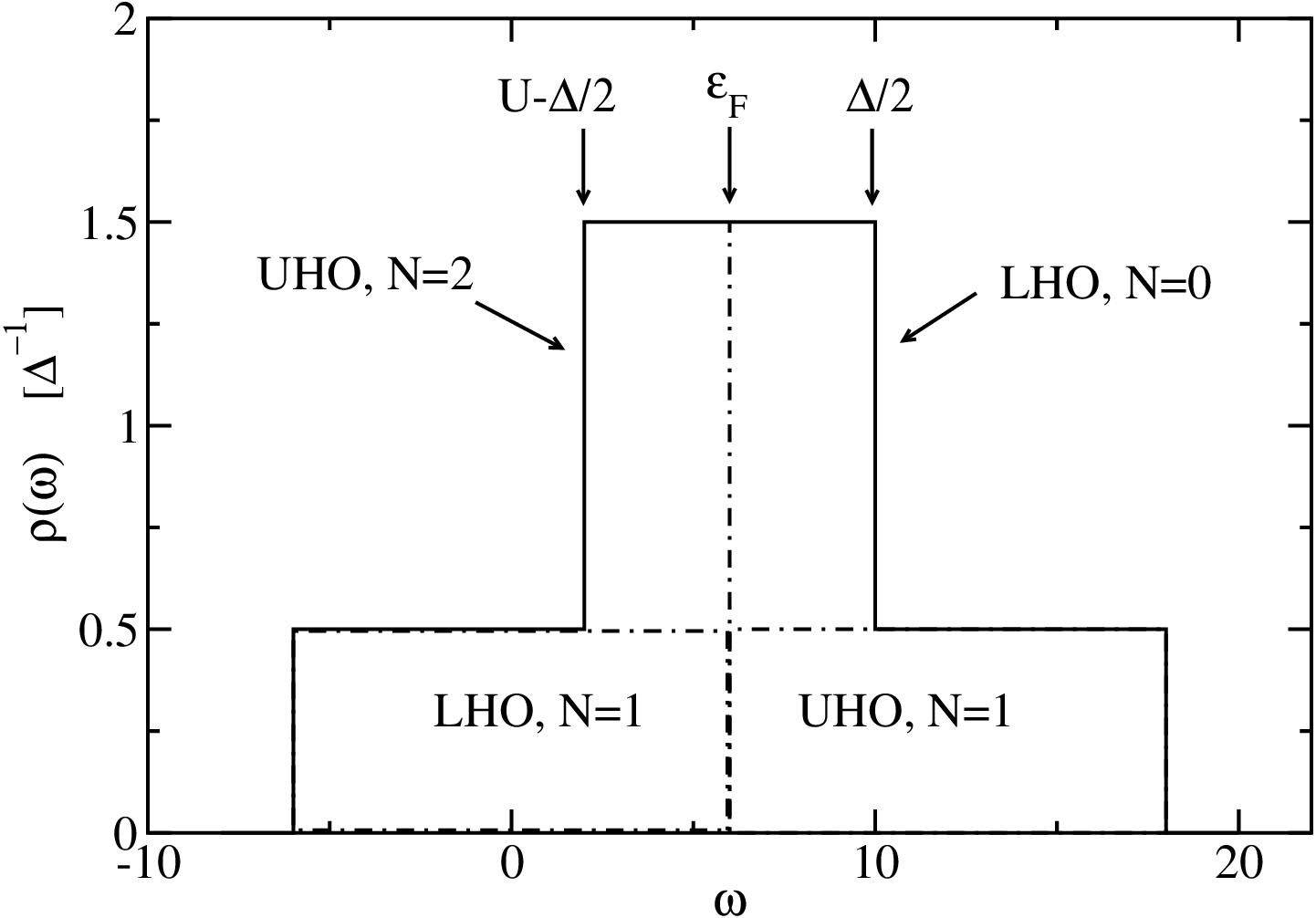}
\caption{Density of states in the atomic limit.  The figure shows different contributions to the ensemble-averaged density of states in the limit $t=0$ for $U=12$ and $\Delta=20$.  The spectral weights contributed by lower Hubbard orbitals (LHO) and upper Hubbard orbitals (UHO) in their different filling states are shown.  For comparison, the noninteracting density of states is $\Delta^{-1}$ for $-\frac 12\Delta < \omega < \frac 12 \Delta$.  The density of states in the ``central plateau'' is $\frac 32 \Delta^{-1}$ and is thus enhanced by interactions, relative to the noninteracting case.  The central plateau extends over $[U-\frac 12 \Delta,\frac 12 \Delta]$ and is the region where the LHO and UHO coexist.}
\label{fig:atomic}
\end{figure}

At half-filling ($\ef=U/2$), the ensemble-averaged density of states is
\begin{eqnarray}
\rho(\omega) &=&\frac 1\Delta \int_{-\Delta/2}^{\Delta/2} d\epsilon \rho_\epsilon(\omega) \nonumber \\
&=& \frac{1}{\Delta} \left [ \Theta\left( \omega - U +\half \Delta \right ) \Theta\left (\half \Delta-\omega \right ) \right . \nonumber \\
&& \left .+\frac 12 \Theta\left( \omega +\half U \right ) \Theta\left (\frac 32U -\omega \right ) 
\right ]. 
\label{eq:atomicDOS} 
\end{eqnarray}
where $\Theta(x)$ is the step function. 
The  result (\ref{eq:atomicDOS}) is illustrated in Fig.~\ref{fig:atomic}.   This figure explicitly shows the spectral weight contributed by the LHO and UHO in their different filling states.  For this work, the most important aspects of the figure are (i) that both LHO and UHO contribute spectral weight at $\ef$ for $\ef \in [U-\half \Delta,\half \Delta]$, and (ii) that for this range of $\ef$ there is a ``central plateau" where interactions enhance $\rho(\ef)$ relative to the noninteracting value $\Delta^{-1}$. 

\subsection{Two-Site Case}
\label{largegap}

\subsubsection{Preliminary Discussion}

\begin{figure}[tb]
\includegraphics[width=\columnwidth]{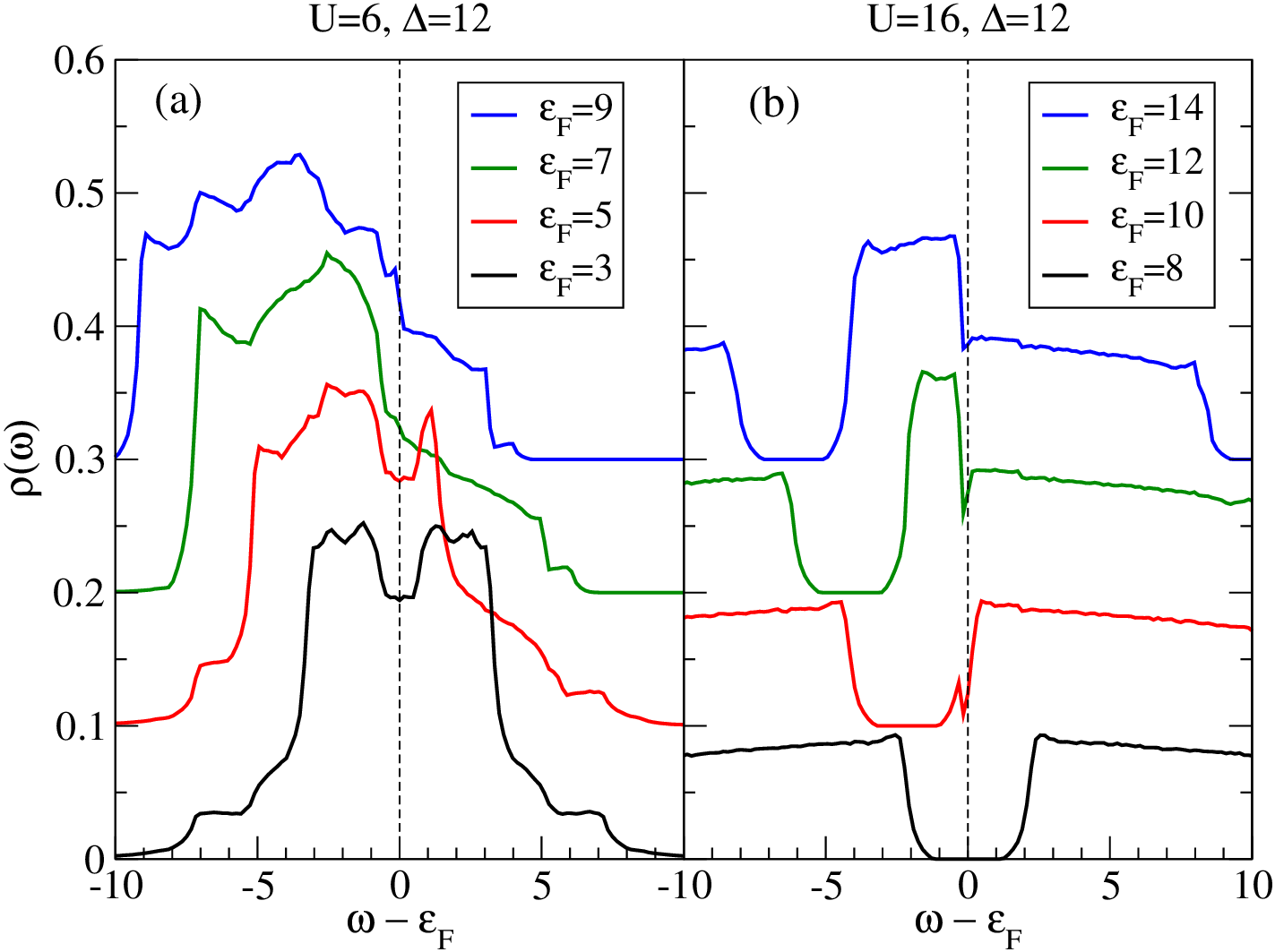}
\caption{(color online) Exact numerical solution for the density of states. Density
  of states for (a) $U < \Delta$ and (b) $U > \Delta$ for different
  values of $\ef$.  In both panels, half-filling corresponds to
  $\ef=U/2$. Curves are offset for clarity. Densities of states are
  averaged over $10^6$ random disorder configurations.  }
\label{fig:numerical}
\end{figure}

The results of exact numerical calculations of the density of states
are shown in Fig.~\ref{fig:numerical} for two cases: $U <\Delta$ and
$U>\Delta$.  We track the evolution of the density of states as a
function of $\ef$ in both cases.  When $U<\Delta$, there is a broad
ZBA centered at $\ef$ for $\ef =3$ and $\ef=5$.  However, the ZBA is
unresolvable when $\ef$ is outside the central plateau.  When
$U>\Delta$, there is a Mott gap at half-filling, and a ZBA forms as
one dopes away from half-filling.  This ZBA is qualitatively different
from that found near half-filling for $U<\Delta$ and has a width of
order $t^2/U$.  In this section, we focus on the large ZBA that appears 
near half-filling for $U<\Delta$.

The approach we take 
is to calculate the density of states $\rho_{\epsilon_1,\epsilon_2}(\omega)$ for a single two-site AHM with site energies $\epsilon_1$ and $\epsilon_2$.  The density of states is then averaged over all possible configurations,
\begin{equation}
\rho(\omega) = \int_{-\frac 12 \Delta}^{\frac 12 \Delta}d \epsilon_1 \int_{-\frac 12 \Delta}^{\frac 12 \Delta} d\epsilon_2 \,\rho_{\epsilon_1,\epsilon_2}(\omega).
\end{equation}

\begin{figure}[tb]
\begin{center}
\includegraphics[width=\columnwidth]{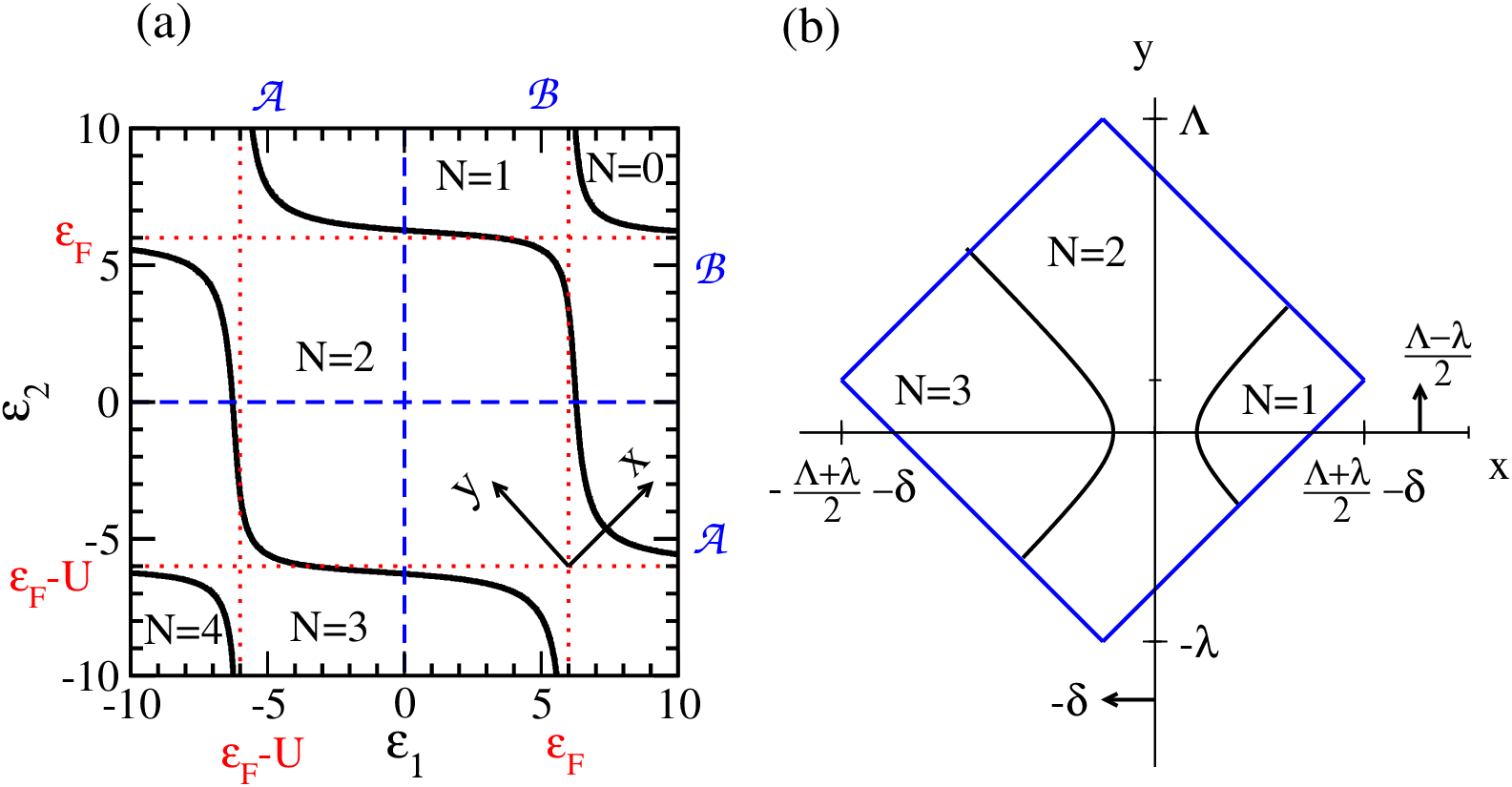}
\end{center}
\caption{(color online) Phase diagram of the two-site AHM.  (a)  Filling states as a function of $\epsilon_1$ and $\epsilon_2$ for $\Delta=20$, $U=12$, $t=1$, and $\varepsilon_F = U/2$, corresponding to half-filling on average.  Also shown are the regions ${\cal A}$ and ${\cal B}$.   Solid black lines indicate phase boundaries for $t$ nonzero;  dotted red lines at $\ef$ and $\ef-U$ indicate phase boundaries for $t=0$; dashed blue lines indicate boundaries of regions ${\cal A}$ and ${\cal B}$. (b) Integration region ${\cal BA}$ for $\ef = \half U + \delta$, shown in terms of transformed coordinates.}
\label{fig:phasediag}
\end{figure}

To simplify the analytic calculations, it is useful to partition the integration range $[-\frac 12 \Delta,\frac 12\Delta]$ into subranges  ${\cal A}=[-\half \Delta,0]$ and  ${\cal B}=[0,\half\Delta]$, as illustrated in Fig.~\ref{fig:phasediag}. Sites whose UHO lies near $\ef$ belong to region $\cal A$, while sites whose LHO lies near $\ef$ belong to region $\cal B$.
We have argued\cite{Hongyi2010a,Wortis2010} that the ZBA comes from level repulsion between LHO and UHO on neighboring sites, and it is indeed suggested by Fig.~\ref{fig:levelsplit}(b) and (c) that the important configurations have
$\epsilon_1 \in \cal A$, $\epsilon_2 \in \cal B$ or $\epsilon_1 \in \cal B$, $\epsilon_2 \in \cal A$.   The simplest approximation is to treat these configurations carefully, while treating the other configurations in the atomic limit.  As we show, this turns out to be sufficient to understand the essential physics of the ZBA.


We denote by $\rho_{\cal XY}(\omega)$ the density of states ensemble-averaged over sites with $\epsilon_1\in \cal X$ and $\epsilon_2\in {\cal Y}$,
\begin{equation}
\rho_{\cal XY}(\omega) = \mbox{Im}\int_{\cal X} d\epsilon_1 \int_{\cal Y} d\epsilon_2 \, \rho_{\epsilon_1,\epsilon_2}(\omega).
\end{equation}
 $\rho_{\cal AA}(\omega)$ and $\rho_{\cal BB}(\omega)$ are evaluated in the atomic limit, using Eq.~(\ref{eq:1sdos}),
\begin{eqnarray}
\rho_{\cal AA + BB}(\omega) &\equiv& \rho_{\cal A A}(\omega)+\rho_{\cal BB}(\omega) \nonumber \\
&=& \frac{1}{2\Delta} \Bigg [\Theta\left (\half \Delta-\omega\right )\Theta\left (\omega-U+\half\Delta\right ) \nonumber  \\
&& + \half \Theta(U-|\omega-\ef|) \Bigg ] 
\end{eqnarray}
For $\omega$ and $\varepsilon_F$ near $U/2$ (half-filling), $\rho_{\cal AA+BB}(\omega) = \frac{3}{4}\Delta^{-1}$. Using $\rho_{\cal BA}(\omega) = \rho_{\cal AB}(\omega)$, 
the total density of states is 
\begin{equation}
\rho(\omega) \approx \frac{3}{4\Delta}+  2\rho_{\cal BA}(\omega).
\label{eq:rhoeq}
\end{equation}
A more careful derivation of $\rho_{\cal AA+BB}(\omega)$ finds
corrections to the atomic limit approximation on the energy scale
$|\omega-\varepsilon_F| < O(t^2/U)$. 

The next step is to evaluate 
\begin{eqnarray*}
\rho_{\cal BA}(\omega) &=& -\frac{1}{\pi\Delta^2} \int_{\cal B}
 d\epsilon_1\int_{A} d\epsilon_2\,
 \mbox{Im} G_{\epsilon_1,\epsilon_2}(\omega),
 \end{eqnarray*}
 with $G_{\epsilon_1,\epsilon_2}(\omega)$ the retarded Green's function averaged over sites and spins.
 It will be convenient to change integration variables to 
 \begin{subequations}
\begin{eqnarray}
x &=& \frac{\epsilon_2 + U+\epsilon_1}{2} - \ef \\
y &=& \frac{\epsilon_2+U-\epsilon_1}{2},
\end{eqnarray}
\label{eq:xycoord}
\end{subequations}
and write
\begin{equation}
\rho_{\cal BA}=-\frac{2}{\pi \Delta^2}  \int_{-\lambda}^{\Lambda} dy
\int_{-x_y-\delta }^{x_y-\delta} dx \, \mbox{Im } G_{x,y}(\omega)
\end{equation}
where the factor of 2 is the Jacobian for the transformation, and the integration
limits are
\begin{equation}
x_y = \frac{\Lambda+\lambda}2 - \left | y - \frac{\Lambda-\lambda}2 \right |  
\label{eq:xy}
\end{equation}
 and
\begin{equation}
\lambda \equiv \frac{\Delta-U}{2}; \quad \Lambda=\frac U2.
\label{eq:lambda}
\end{equation} 
 The Fermi energy is written
\begin{equation}
\ef = \frac{U}{2} +\delta.
\label{eq:delta}
\end{equation}
This equation defines $\delta$ such that half-filling corresponds to $\delta=0$.  
Figure~\ref{fig:phasediag} illustrates the integration region and gives the graphical meaning of $\lambda$, $\Lambda$, and $\delta$.

 The phase diagram Fig.~\ref{fig:phasediag} shows that there are three filling states in ${\cal BA}$, with $N=1,2$, or 3 electrons.  We now find the ground state wavefunctions, energies, and phase boundaries for the different filling states.

\subsubsection{Ground states in region ${\cal BA}$}
We will determine the ground state wavefunction in the region $\cal BA$ using a truncated basis set that discards high energy states.  These high energy states modify the ground state wavefunctions and energies by $O(t^2/U)$, and our approximation is consequently valid for $U\gg t$.  

In $\cal BA$, the one-electron ground state in the atomic limit is $|0\uparrow\rangle$  because $\epsilon_1 > \epsilon_2$.   Making $t$ nonzero mixes in a small amount of $|\uparrow 0\rangle$, proportional to $t^2/(\epsilon_1-\epsilon_2)$.  However, in $\cal BA$, $\epsilon_1\sim \epsilon_2+U$, so the mixing is of order $t^2/U$ and is neglected in our approximation.
The one-electron ground state is thus 
\begin{equation}
|G1\rangle  \approx |0\uparrow\rangle.
\end{equation}
 Similarly, the three-electron ground state is $|G3\rangle \approx |\uparrow 2\rangle$.

The two electron ground state is found by diagonalizing the AHM Hamiltonian in the reduced basis $\{|s\rangle,|02\rangle\}$,  where  
\begin{eqnarray}
|s\rangle &=& \frac{1}{\sqrt{2}} \left ( |\uparrow \downarrow\rangle - |\downarrow\uparrow\rangle \right ) 
\end{eqnarray}
is the singlet state.  For $\epsilon_1\sim\epsilon_2+U\sim\ef$, each of these basis states has an energy $\sim 2\ef-U$.  The discarded basis state $|20\rangle$ has an energy $\sim 2\ef+U$, and the amount of $|20\rangle$ mixed into the ground state by $t$ is therefore $\sim O(t^2/U)$, which we ignore.
  The Hamiltonian matrix in the reduced basis is
\begin{equation}
H = \left[ \begin{array}{cc}
\epsilon_1+\epsilon_2 & -\sqrt{2} t \\
-\sqrt{2} t & 2\epsilon_2 +U 
\end{array} \right ]
\end{equation}
which has eigenenergies 
\begin{equation}
E_\pm = \frac{\epsilon_1+3\epsilon_2+U}{2} \pm \sqrt{\left ( \frac{\epsilon_2+U-\epsilon_1}{2}\right )^2 + 2t^2}
\label{eq:2e}
\end{equation}
and eigenstates
\begin{equation}
|\pm\rangle = \alpha_{1\pm}|s\rangle + \alpha_{2\pm}|02\rangle
\label{eq:pm}
\end{equation}
\begin{equation}
\alpha_{1\pm}^2 = \frac{ \sqrt{y^2 + 2t^2} \mp y}{2 
\sqrt{y^2 + 2t^2} }; \quad \alpha_{2\pm}^2 = 1-\alpha_{1\pm}^2
\end{equation}
where  $y$ is defined in (\ref{eq:xycoord}).  The two-electron ground state is 
$|G2\rangle = |-\rangle$.
The different ground states, their energies, and the phase boundaries between them are tabulated in Table~\ref{tab:gs}.  The next step is to calculate the density of states for each filling state.

\begin{table}[tb]
\caption{Approximate $N$-electron ground states $|GN\rangle$ and their energies $E_G$ for the two site model with $\Delta>U\gg t$ in the region $\cal BA$. Variables $x$ and $y$ are defined in
(\protect\ref{eq:xycoord}), and $E_-$ is defined in (\protect\ref{eq:2e}).}  
\label{tab:gs}
\begin{tabular}{c|c|c|c}
$N$ & $E_G-\ef N$ & $|GN\rangle$ & Ground state when \\
\hline 
1 & $\epsilon_2-\ef$ & $|0\uparrow\rangle$ & $x > \sqrt{y^2+2t^2}$ \\
2 & $E_- - 2\ef$ & $\alpha_{1-}|s\rangle + \alpha_{2-} |02\rangle$ & $| x| < \sqrt{y^2+2t^2}$\\
3 & $\epsilon_1 + 2\epsilon_2 + U -3\ef$ & $|\uparrow 2\rangle$ & $x < -\sqrt{y^2+2t^2}$\\
\hline
\end{tabular}
\end{table}

\subsubsection{Density of states for the three-electron ground state}
First, we calculate the contribution to the density of states from the 3-electron ground state.  Throughout this work,
we keep only terms with poles near $\ef$, meaning that terms with poles near $\epsilon_1+U$ or $\epsilon_2$ are discarded.  The spin- and site-averaged Green's function is then
\begin{eqnarray}
G^{3e}_{\epsilon_1,\epsilon_2} (\omega)&\approx& \frac 14 \left \{ \sum_\pm \frac{
|\langle \pm | c_{1\uparrow} |\uparrow 2 \rangle|^2+|\langle \pm | c_{2\uparrow} |\uparrow 2 \rangle|^2
}{\omega^+ - (\epsilon_1+2\epsilon_2+U) + E_\pm} \right . \nonumber \\
&& \left . +\frac{|\langle t | c_{2\uparrow} |\uparrow 2\rangle|^2
+|\langle \uparrow\uparrow | c_{2\downarrow} |\uparrow 2 \rangle|^2
}{\omega^+-(\epsilon_1+2\epsilon_2+U) + (\epsilon_1+\epsilon_2)} \right \} ,
\label{eq:g3e1}
\end{eqnarray}
where
\begin{equation}
|t\rangle = \frac{1}{\sqrt{2}} \left ( |\uparrow \downarrow\rangle + |\downarrow\uparrow\rangle \right )
\end{equation}
is a triplet state. 
Using $\alpha_{2\pm}^2 = 1-\alpha_{1\pm}^2$, we reduce Eq.~(\ref{eq:g3e1}) to
\begin{equation}
G^{3e}_{\epsilon_1,\epsilon_2}(\omega) = \frac 14 \sum_\pm \frac{1-\alpha_{1\pm}^2/2}{\tilde\omega^+ - x \pm \sqrt{y^2+2t^2}}  + \frac 38 \frac{1}{\tilde\omega^+-x-y}
\end{equation}
where $x$ and $y$ are defined in (\ref{eq:xycoord}) and $\tilde\omega = \omega-\ef$. 

The ground state has three electrons for 
\begin{equation}
-x_y-\delta < x < -\sqrt{y^2+2t^2},
\end{equation}
where
the upper limit is the phase boundary between two- and three-electron states (c.f.\ Table~\ref{tab:gs}), and the lower limit [c.f.\ Eq.~(\ref{eq:xy})] is the boundary of region $\cal BA$.
Then 
\begin{eqnarray}
\rho_{\cal BA}^\mathrm{3e} &=& \frac{1}{2\Delta^2}  \int_{-\frac {\lambda+\delta} 2}^{\frac {\Lambda+\delta} 2} dy
\int_{-x_y-\delta}^{-\sqrt{y^2+2t^2}} dx  \, \nonumber \\
&& \times
\Big [ (1-\alpha_{1+}^2/2)\delta(\tilde\omega-x+\sqrt{y^2+2t^2})  \nonumber \\
&& 
+ (1-\alpha_{1-}^2/2 )\delta(\tilde\omega-x-\sqrt{y^2+2t^2}) \nonumber \\
&& +\frac 32 \delta(\tilde\omega-x-y) \Big ]. 
\end{eqnarray}

The integration over $x$ is straightforward because of the delta-functions, which introduce the constraints $\tilde \omega<0$ and 
\begin{eqnarray*}
 |y|<\frac{|\tilde\omega|}{2}\mbox{Re }\sqrt{1-\frac{8t^2}{\tilde\omega^2}},&&\quad \mbox{first term,} \\
- \frac{\lambda+\delta-|\tilde\omega|}{2}<y<\frac{\Lambda+\delta-|\tilde\omega|}{2},&&\quad \mbox{second term,} \\
 -\left( \frac{|\tilde\omega|}{2} - \frac{t^2}{|\tilde\omega|} \right ) <y<\frac{\Lambda+\delta-|\tilde\omega|}{2},&&\quad \mbox{third term.}
\end{eqnarray*}
The result for the first term is valid for $0>\tilde\omega > -\lambda - \delta$, i.e.\ for $\omega<\ef$ and in the central plateau.  In deriving these results, we have neglected terms of order $t^2/\lambda$ and $t^2/\Lambda$.
We now integrate over $y$ using
\begin{equation}
\int dy\,\left ( 1-\frac{\alpha_{1\pm}^2}{2}\right ) = \frac 14 \left ( 3y \pm \sqrt{y^2+2t^2}\right ),
\end{equation}
 to get the three electron contribution to the density of states,
\begin{eqnarray}
\rho_{\cal BA}^\mathrm{3e}(\tilde\omega) &=& \frac{3\Theta(-\tilde\omega)}{8\Delta^2} 
\left [  \frac{2\lambda+\Lambda}{3} +\delta 
- {|\tilde\omega|}\left ( 1- \mbox{Re } \sqrt{1-\frac{8t^2}{\tilde\omega^2}} \right )   \right . \nonumber \\
&&+\left. \Theta \left (-\tilde\omega-\frac{2t^2}{\Lambda+\delta}\right )
\left ( \Lambda+\delta - \frac{2t^2}{|\tilde\omega|} \right ) \right ].
\label{eq:3edos}
\end{eqnarray}
To simplify the final expression, we have taken $\sqrt{\lambda^2+2t^2} \approx \lambda$ and $\sqrt{\Lambda^2+2t^2}\approx \Lambda$.

\subsubsection{Density of states for the one-electron ground state}
The derivation of the one-electron contribution to the density of states parallels that of the three-electron contribution.  The Green's function is
\begin{eqnarray}
G_{\epsilon_1,\epsilon_2}^{1e}(\omega)& \approx&\frac 14 \left \{  \sum_\pm
\frac{|\langle \pm | c^\dagger_{1\downarrow} |0\uparrow \rangle|^2
+|\langle \pm | c^\dagger_{2\downarrow} |0\uparrow \rangle|^2
}{\omega^+ + \epsilon_2 -E_\pm } \right .\nonumber \\
&& \left . +
\frac{ |\langle t | c^\dagger_{1\downarrow} |0\uparrow \rangle|^2
+
|\langle \uparrow\uparrow | c^\dagger_{1\uparrow} |0\uparrow \rangle|^2
}{\omega^+ + \epsilon_2 - (\epsilon_1+\epsilon_2)}  \right \} \nonumber \\
& = &\frac 14 \sum_\pm \frac{1-\alpha_{1\pm}^2/2}{\tilde\omega^+ - x \mp \sqrt{y^2+2t^2}}  + \frac 38 \frac{1}{\tilde\omega^+-x+y}. \nonumber \\
\end{eqnarray}
The integration region is $x_y-\delta > x > \sqrt{y^2+2t^2}$ with $x_y$ given by (\ref{eq:xy})
and
\begin{eqnarray*}
&&\rho_{\cal BA}^\mathrm{1e} = \frac{1}{2\Delta^2}  \int_{-\frac {\lambda-\delta} 2}^{\frac {\Lambda-\delta} 2} dy
\int_{\sqrt{y^2+2t^2}}^{x_y-\delta} dx  \, \\
&& \times
 \Big [ (1-\alpha_{1+}^2/2) \delta(\tilde\omega-x-\sqrt{y^2+2t^2}) \\
&&
+ (1-\alpha_{1-}^2/2) \delta(\tilde\omega-x+\sqrt{y^2+2t^2}) +\frac 32 \delta(\tilde\omega-x+y) \Big ]  
%
\end{eqnarray*}
Letting $x\rightarrow -x$, this is the same as 
$\rho_{\cal BA}^\mathrm{3e}(-\tilde\omega)$ for $\delta\rightarrow -\delta$.  Thus
\begin{eqnarray}
\rho_{\cal BA}^\mathrm{1e}(\tilde\omega) &&= \frac{3\Theta(\tilde\omega)}{8\Delta^2} 
\Bigg [  \frac{2\lambda+\Lambda}{3}-\delta - {\tilde\omega}\left ( 1- \mbox{Re } \sqrt{1-\frac{8t^2}{\tilde\omega^2}} \right )
\nonumber \\
&&+ \Theta \left (\tilde\omega-\frac{2t^2}{\Lambda-\delta}\right )
\left ( \Lambda-\delta - \frac{2t^2}{\tilde\omega} \right ) \Bigg ].
\end{eqnarray}

\subsubsection{Density of states for the two-electron  ground state}
Finally, the Green's function for the two-electron ground state is
\begin{eqnarray}
G_{\epsilon_1,\epsilon_2}^{2e} &\approx& \frac 12 \left \{ 
\frac{|\langle 0\downarrow | c_{1\uparrow} |G2\rangle|^2
+|\langle 0\downarrow | c_{2\uparrow} |G2\rangle|^2}
{\omega^+ - E_- + \epsilon_2} \right . \nonumber \\
&& \left . + \frac{|\langle \uparrow 2 | c^\dagger_{1\uparrow} |G2\rangle|^2
+|\langle \uparrow 2 | c^\dagger_{2\uparrow} |G2\rangle|^2}{\omega^+ + E_- - (\epsilon_1 + 2\epsilon_2+U)} \right \}
\label{eq:2eG} \\
&=&  \frac 12 \sum_\pm \frac{1-\alpha_{1-}^2/2}{\tilde\omega^+ - x\pm\sqrt{y^2+2t^2}} 
\end{eqnarray}
Then,  
\begin{eqnarray}
&& \rho_{\cal BA}^\mathrm{2e}=\frac{1}{\Delta^2}  \int_{-\lambda+|\delta|}^{\Lambda-|\delta|} dy
\int_{x_1}^{x_2} dx \,\left (1-\frac{\alpha_{1-}^2}{2}\right ) \nonumber \\
&&\times  \sum_\pm
 \delta\left( \tilde\omega -x \pm \sqrt{y^2+2t^2}\right)  
\end{eqnarray}
where 
$x_1 = -\min(x_y+\delta,\sqrt{y^2+2t^2})$, 
$x_2 = \min(x_y-\delta,\sqrt{y^2+2t^2})$.
Performing the integrations over $x$ and $y$ gives
\begin{eqnarray}
&&\rho_{\cal BA}^\mathrm{2e}(\tilde\omega)=\frac{3}{4\Delta^2}\left[\frac{2\lambda+ \Lambda}3 + |\tilde\omega|\left ( 1-\mbox{Re }\sqrt{1-\frac{8t^2}{\tilde\omega^2}}\right ) \right] \nonumber \\
&& + \frac{3\delta}{4\Delta^2}[\Theta(\tilde\omega)-\Theta(-\tilde\omega)]
\end{eqnarray}
%

\subsubsection{Total density of states}
\begin{figure}
\includegraphics[width=\columnwidth]{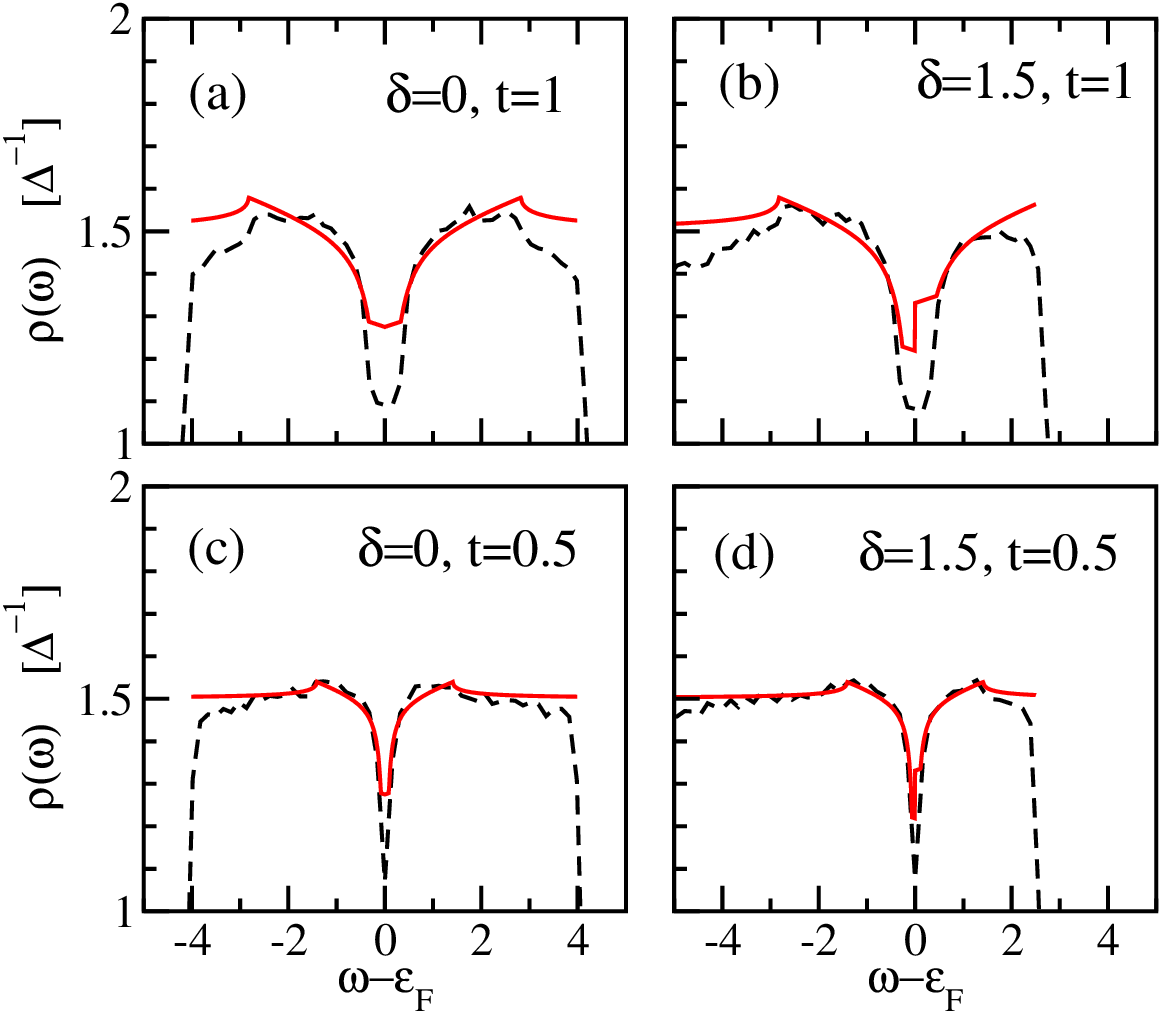}
\caption{(color online) Density of states for the two-site ensemble-averaged Anderson-Hubbard model
for different values of $t$ and $\delta$.  Results are shown for exact numerical solution of the AHM (dashed black curves) and for the approximate result, Eq.~(\protect\ref{eq:finaldos}) (solid red curves).  Model parameters are $U=12$, $\Delta=20$.
  Recall that $\ef = U/2 + \delta$ and that $\delta=0$ corresponds to half-filling.}
\label{fig:ED1}
\end{figure}

Putting the results of the different calculations together, we arrive at our final result for the total density of states (valid in the central plateau)
\begin{eqnarray} 
\rho(\omega) &=& \rho_\mathrm{\cal AA+BB}(\omega) + 2\rho_{\cal BA}^\mathrm{1e}(\omega) +2\rho_{\cal BA}^\mathrm{2e}(\omega) + 2\rho_{\cal BA}^\mathrm{3e}(\omega)  \nonumber \\
&& = \frac{3}{2\Delta} + \frac{3}{2\Delta^2} \Bigg [ - \frac U 4 + 
\frac{\delta_s}{2} \nonumber \\
&& +\frac{|\tilde\omega|}{2} 
\left ( 1- \mbox{Re } \sqrt{1-\frac{8t^2}{\tilde\omega^2}} \right )  \nonumber \\
&&+\left .  \Theta \left (|\tilde\omega|-\frac{4t^2}{U-2\delta_s}\right ) 
\left ( \frac{U-2\delta_s}{4} - \frac{t^2}{|\tilde\omega|} \right ) \right ]
\label{eq:finaldos}
\end{eqnarray}
where $\delta_s = \delta\sgn(\tilde\omega)$ and $\tilde\omega=\omega-\ef$.  The parameters $\lambda$, $\Lambda$ and $\delta$ are defined in Eqs.~(\ref{eq:lambda}) and (\ref{eq:delta}).

As a check of Eq.~(\ref{eq:finaldos}), we let $t\rightarrow 0$, in which case
\[
\rho(\omega) \rightarrow  \frac{3}{2\Delta}
\]
which agrees with the previous atomic limit calculation for the central plateau.
Equation (\ref{eq:finaldos}) is plotted in Fig.~\ref{fig:ED1} in comparison with the exact density of states determined from numerically diagonalizing the AHM.

\section{Discussion}
\label{discussion}

Figure~\ref{fig:ED1} compares Eq.~(\ref{eq:finaldos}) to exact disorder-averaged numerical calculations for the density of states.  The theory works well for $|\tilde \omega| > 4t^2/U$, out to the edges of the central plateau where it breaks down  (for example, near $\tilde \omega = \pm 4$ in Fig.~\ref{fig:ED1}(a)).   The theory neglects terms of order $t^2/U$, and therefore fits the numerics better when $t$ is smaller, as shown in Fig.~\ref{fig:ED1}(c) and (d).  The fit for $|\tilde \omega|<4t^2/U$ is not especially good, but can be improved significantly by considering corrections of order $t^2/U$ that were neglected in the previous section; we haven't included these corrections because they complicate $\rho(E)$ significantly without adding physical insight.  The focus of this discussion is therefore $|\tilde \omega| > 4t^2/U$.

The main qualitative idea that we emphasize in this section is that there are two distinct physical mechanisms that lead to the ZBA in Eq.~(\ref{eq:finaldos}).  Both mechanisms occur for configurations where the LHO of one site and the UHO of the other site are nearly degenerate with $\ef$, namely for $\epsilon_1 \sim \epsilon_2+U \sim \varepsilon_F$ or $\epsilon_2 \sim \epsilon_1+U \sim \varepsilon_F$.   The first mechanism is similar to that outlined in the mean-field calculation in Sec.~\ref{HF}: level repulsion, caused by hybridization of many-body states, shifts states away from $\ef$.  The second mechanism does not have a mean-field counterpart:  level repulsion gaps the spectrum of low energy triplet excitations.

 The first mechanism underlies the second last term in Eq.~(\ref{eq:finaldos}),
\begin{equation}
\frac{3|\tilde \omega|}{4\Delta^2}\left ( 1- \mbox{Re }\sqrt{1-\frac{8t^2}{\tilde \omega^2}} \right ).
\label{eq:2t}
\end{equation}
This term rises linearly from $\tilde \omega=0$ and is peaked at $\tilde \omega= \pm 2\sqrt{2}t$, which defines the width of the ZBA in Fig.~\ref{fig:ED1}.  
In our calculations, this term comes from transitions between two-electron
singlet states and states with one or three electrons, and it is the level repulsion between the two-electron states that causes the ZBA.  In the case, for example,  where $\epsilon_1$  and $\epsilon_2+U$ lie near $\varepsilon_F$, there are two nearly-degenerate two-electron singlets, $|s\rangle$ and $|02\rangle$; these hybridize as a result of the matrix element $t$ to form bonding and antibonding many-body states with energies (from Eq.~(\ref{eq:2e})),
\begin{equation}
E_\pm \approx 2\ef - U \pm \sqrt{2}t.
\end{equation}
Thus, the level repulsion between $|s\rangle$ and $|02\rangle$ shifts the many-body orbital energies up or down by $O(t)$.  Starting from the two-electron ground state, with energy $E_-$, one has transitions
\begin{equation}
\begin{array}{l}
\alpha_{1-}|s\rangle + \alpha_{2-}|02\rangle
\stackrel{c_{1\sigma}^\dagger, c_{2\sigma}^\dagger}{\rightarrow} |\sigma 2\rangle, \\
\alpha_{1-}|s\rangle + \alpha_{2-}|02\rangle
\stackrel{c_{1\sigma}, c_{2\sigma}}{\rightarrow} |0 \sigmab\rangle.
\end{array}
\label{eq:3exa}
\end{equation}
We showed in Sec.~\ref{largegap} that the three-electron energy is $\epsilon_1+2\epsilon_2+U$ (which is approximately $3\ef-U$), and the one-electron
energy is $\epsilon_2 $ (approximately $\ef-U$), so that the transition energies in Eq.~(\ref{eq:3exa}) are
\begin{equation}
\omega_\pm \approx \ef \pm \sqrt{2}t.
\end{equation}
Because $\omega_\pm$ are shifted by $O(t)$ away from $\ef$, the density of states at $\ef$ is reduced as $t$ increases.  As indicated above, this mechanism for depleting the low energy density of states is similar to the mean-field mechanism discussed in Sec.~\ref{HF}, where level repulsion between molecular states on opposite sides of $\varepsilon_F$ increases the  energy required to add or remove an electron.  In this sense, the  second-last term in (\ref{eq:finaldos}) is  Altshuler-Aronov-like.

The second mechanism does not have a mean-field counterpart, and results in the last term in Eq.~(\ref{eq:finaldos}) 
\begin{equation}
\frac{3}{2\Delta^2} \Theta \left (|\tilde\omega|-\frac{4t^2}{U-2\delta_s}\right ) 
\left ( \frac{U-2\delta_s}{4} - \frac{t^2}{|\tilde\omega|} \right ).
\label{eq:t3}
\end{equation}
This term varies as $|\tilde \omega|^{-1}$ down to the low energy cutoff at $|\tilde \omega| \sim 4t^2/U$,  and makes the dominant contribution to the shape of the ZBA. The cutoff comes from the boundary between the region ${\cal BA}$ and the region ${\cal BB}$ in Fig.~\ref{fig:phasediag}, where the approximate one- and two-electron wavefunctions used in deriving $\rho(E)$ cease to be valid.
   
   In our calculations, Eq.~(\ref{eq:t3}) comes
from transitions between one- or three-electron ground states, and two-electron triplet excitations.  For the three-electron ground state, for example, these excitations have
the form
\begin{equation}
|\uparrow 2\rangle \stackrel{c_{2\downarrow}}{\rightarrow} |\uparrow\uparrow\rangle, \quad
|\uparrow 2\rangle \stackrel{c_{2\uparrow}}{\rightarrow} |t\rangle.
\label{eq:3ex}
\end{equation}
As mentioned above, the three-electron energies are nearly independent of $t$; the triplet energies are also independent of $t$, however, so that the transition energies are not shifted by level repulsion.  The mechanism for depleting the low energy density of states in this term is therefore {\em not} that of Altshuler and Aronov.

Instead, it is the fact that a gap in the triplet spectrum opens as
$t$ increases that causes a depletion of states near $\varepsilon_F$
(this gap was illustrated in Fig.~\ref{fig:levelsplit}).  This gap
occurs for configurations of $(\epsilon_1, \epsilon_2)$ that have one-
or three-electron ground states in the atomic limit, but two-electron
ground states when $t$ is nonzero.  For example, when $\epsilon_1$ and
$\epsilon_2+U$ both lie slightly below $\varepsilon_F$, the
atomic-limit ground state has three electrons and triplet excitations
as in (\ref{eq:3ex}) are possible.  When $t$ is nonzero, the
two-electron ground state energy $E_-$ is reduced by $O(t)$, while the
three electron ground state energy is reduced by $O(t^2/U)$.  For
sufficiently large $t$, the two-electron ground state has the lower
energy and the triplet excitation is eliminated (i.e.\ the only
possible tunneling processes are to one- or three-electron final
states).  In summary, the ZBA in the final term of
Eq.~(\ref{eq:finaldos}) occurs because the phase space for low energy
triplet excitations is reduced when $t$ increases.

As we discus elsewhere,\cite{Hongyi2010a,Wortis2010} this calculation
sheds light on the empirical observation, made for larger systems,
that the width of the ZBA is of order $t$.\cite{Chiesa2008} A na\"ive
argument based on the disorder-free Hubbard model would suggest that
the ZBA might have a conventional Altshuler-Aronov form, but with an
effective exchange interaction $V_\mathrm{eff}=4t^2/U$, so that the
ZBA should grow with increasing $t^2/U$.  As we have said above, there
are contributions to the density of states of this type; however, we
have just shown that a much larger effect, of order $t$, comes from
configurations with the LHO and UHO on neighboring sites degenerate.

We note that this explanation appears to contradict numerical evidence
from the work of Chiesa et al.\cite{Chiesa2008} on two-dimensional
clusters that a large ZBA persists far from half-filling and for large
$U$, since configurations with a degenerate LHO and UHO do not occur
in these cases; as we have shown in Fig.~\ref{fig:numerical},  the ZBA
vanishes rapidly (with increasing disorder) in the two-site model when $\ef$ is outside the central plateau.  To
check this, we have performed preliminary exact diagonalization
calculations for larger clusters (up to 12 sites).  These calculations
find that the width of the ZBA is not linear in $t$ when $\ef$ is
outside the central plateau, and suggest that the physics of the ZBA
changes far from half-filling.  A more detailed study of how the ZBA
evolves with doping needs to be undertaken.

\section{Conclusions}
\label{conclusions}
In summary, we have found that the zero bias anomaly in the two-site Anderson-Hubbard model is the result of strong orbital hybridization in the two-electron ground state for configurations with $\epsilon_1\sim \epsilon_2+U\sim \ef$  or with $\epsilon_2\sim \epsilon_1+U\sim \ef$.  Unlike in the conventional Hubbard model, this hybridization is not suppressed by the on-site interaction $U$, and leads to a level repulsion between molecular orbital energies of order $t$, rather than $t^2/U$.   

The mechanism for the suppression of the tunneling density of states is, at least in part, different from in conventional mean-field models of interacting electrons.  In mean-field theories, interactions cause a shift of molecular orbital energies away from $\ef$ that leads directly to an increase in the energy required to remove or add an electron.  This also occurs in the Anderson-Hubbard model; however, there is an additional depletion of low energy spectral weight because the low energy triplet excitation spectrum is gapped as a result of orbital hybridization.  This mechanism is physically different from that of  Altshuler and Aronov.   

\section*{Acknowledgments}
We thank R.\ Wortis for helpful discussions.  We acknowledge the
support of NSERC of Canada.  H.-Y.C.\ is supported by grant NSC 98-2112-M-003-009-MY3.


\end{document}